# Tunnel Magnetoresistance with Atomically Thin Two-Dimensional Hexagonal Boron Nitride Barriers

*Andre Dankert[1*], M. Venkata Kamalakar[1], Abdul Wajid[1], R. S. Patel[2#], Saroj P. Dash[1†]*

[1]Department of Microtechnology and Nanoscience, Chalmers University of Technology, SE-41296, Göteborg, Sweden

[2]Department of Physics, Birla Institute of Technology and Science Pilani – K K Birla Goa Campus, Zuarinagar– 403726, Goa, India

[*]andre.dankert@chalmers.se; [#]rsp@goa.bits-pilani.ac.in, [†]saroj.dash@chalmers.se

**ABSTRACT**

The two-dimensional atomically thin insulator hexagonal boron nitride (h-BN) constitutes a new paradigm in tunnel based devices. A large band gap along with its atomically flat nature without dangling bonds or interface trap states makes it an ideal candidate for tunnel spin transport in spintronic devices. Here, we demonstrate the tunneling of spin-polarized electrons through large area monolayer h-BN prepared by chemical vapor deposition in magnetic tunnel junctions. In ferromagnet/h-BN/ferromagnet heterostructures fabricated over a chip scale, we show tunnel magneto resistance at room temperature. Measurements at different bias voltages and on multiple devices with different ferromagnetic electrodes establish the spin polarized tunneling using h-BN barriers. These results open the way for integration of 2D monolayer insulating barriers in active spintronic devices and circuits operating at ambient temperature, and for further exploration of their properties and prospects.

## 1. Introduction

The quantum phenomenon of electron tunneling enables novel spintronic, electronic, optoelectronic and superconducting nanodevices with enhanced efficiency and low power consumption [1]. Such tunnel devices typically require growth of insulating materials of few atomic layers thin, which is a major challenge in materials science. Mostly used conventional tunnel barrier materials are made up of metal oxides, which have non-uniform thicknesses, pinholes, defects and trapped charges. These issues compromise the performance and reliability of the devices. Tunnel barriers are main building blocks of magnetic tunnel junctions (MTJs), where two ferromagnetic (FM) contacts are separated by ultra-thin oxide barriers [2-6]. Such MTJs are currently used in read-heads of hard drives and new emerging technologies including magnetic random access memory and spin-transfer torque devices [3-5]. For many of these applications, it is crucial to achieve tunnel magnetoresistance (TMR) signals with optimal junction resistances [3], where the demand for controlling thickness of metal oxide tunnel barriers with atomic level precession pose a serious challenge.

The discovery of the two-dimensional (2D) atomic crystals has opened up the possibility of exploring their fascinating properties [7, 8]. Recently, graphene has been explored as a barrier for spin transport, where TMR and spin filtering effects have been observed [9-19], including the bias dependence of the TMR [20]. However, the absence of a bandgap in semi-metallic and low resistive graphene, leads to an increasing demand for an insulating 2D crystal [7-9]. Hexagonal boron nitride (h-BN) is an insulating isomorph of graphene with a large bandgap of ~ 6 eV, making it an ideal candidate for dielectric substrates and tunnel barriers [8, 21-25]. The atomically thin and inert nature of h-BN can provide the ultimate control over the morphology and can minimize defects related to interface states and interfacial alloy formation. It has been theoretically proposed to use h-BN as tunnel barrier in MTJs to achieve large magnetoresistance signals [26-28]. The spin filtering nature of h-BN/ferromagnet interfaces and tailoring of TMR by uniaxial strain has been proposed [26-28]. Experimentally, h-BN has been used as tunnel barrier for spin injection in lateral graphene spin transport devices [29-31]. The chemical vapor deposited (CVD) h-BN is found to show reproducible tunneling behavior circumventing the conductivity mismatch problem between the ferromagnet and graphene for efficient spin injection [30]. In order to further establish the potential of atomically thin h-BN tunnel barriers, it is important to demonstrate TMR in technologically relevant MTJs, which has not been realized so far.

Here, we employ large area CVD grown monolayer h-BN as a tunnel barrier in MTJs having ferromagnetic $Ni_{80}Fe_{20}$ and Co contacts in a vertical structure. We show that atomically thin CVD h-BN exhibits quantum tunneling of spin polarized electrons for the demonstration of TMR at room temperature. Measurements on multiple devices with different ferromagnetic electrodes and bias dependence of the TMR signal establish the spin polarized tunneling effects using h-BN barriers. Our results demonstrate the integration of atomically thin 2D insulating tunnel barriers in magnetic tunnel devices and the observation of tunnel magnetoresistance at ambient temperatures.

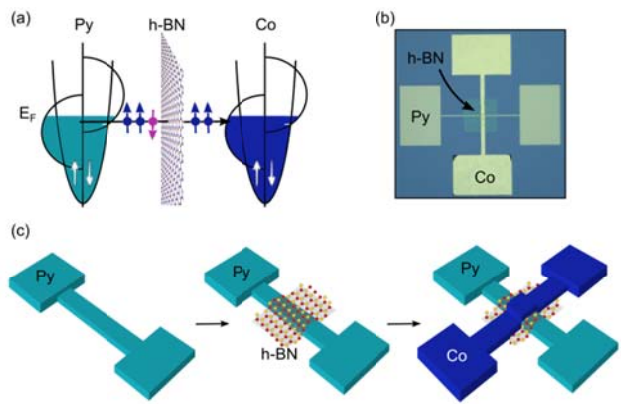

*Figure 1| Hexagonal boron nitride magnetic tunnel junction. (a) Schematics of spin polarized tunneling in a FM/h-BN/FM tunnel junction. (b) Optical micrograph of a fabricated magnetic tunnel junction with Py/h-BN/Co contact. (c) Fabrication steps of magnetic tunnel junction with CVD h-BN and FM electrodes.*

2. Experimental

Magnetic tunnel junctions incorporating a h-BN tunnel barrier between two FM metal electrodes are schematically presented in Fig 1a. When the magnetizations of the FMs are in a parallel orientation it is preferable that spin polarized electrons will tunnel through the h-BN than if they are antiparallel, giving rise to a change in the resistance across the junction [32]. We fabricated magnetic tunnel devices using a h-BN barrier and ferromagnetic $Ni_{80}Fe_{20}$ (Py) and Co electrodes (Fig. 1b). The main fabrication steps of our devices are presented in Fig. 1c. The bottom Co or Py electrodes are prepared on $Si/SiO_2$ substrate by photo lithography, electron beam evaporation and lift off methods. Before deposition of FM electrodes, we deposited a thin layer of Ti in order to increase adhesion of FM to the $SiO_2$ substrate. Subsequently we transferred the CVD grown layer of h-BN on bottom FM electrodes. We have been able to achieve a ripple free transfer of CVD h-BN over large areas on our chips by using a frame assisted process. The CVD h-BN layer used in our experiment was grown on copper substrate (purchased from Graphene supermarket). The h-BN surface is first covered with PMMA, and then isolated from the Cu substrate by etching in a $H_2O_2$-HCl solution. This Cu etchant produces very clean h-BN in comparison to other available chemical etching procedures. The h-BN/PMMA layer is washed with deionized water, and subsequently transferred onto the chip containing bottom FM electrodes in isopropyl alcohol medium. We avoided the use of DI water on FM contacts to minimize the oxidation. However, oxidation and contamination of bottom FM electrode due to air exposure could not be avoided in the present process. After drying the chip in an ambient environment, we annealed it at 150°C for 10min. It has been observed that this annealing step improved adhesion of h-BN with bottom FM electrode. The chip was then washed with acetone to remove the PMMA resulting in chip containing h-BN/FM heterostructures. The top electrodes of ferromagnetic Co and capping Au layer were prepared on the h-BN/FM heterostructures by photo lithography, e-beam evaporation and lift off techniques. The atomic force microscopy was performed with a Bruker Dimension 3100 SPM using tapping mode. The Raman spectrum was measured with a Horiba XploRA system using a 638 nm laser and a grating of 1800 lines/mm. TMR measurements were performed at room temperature with application of an in-plane magnetic field $B_{in}$. We sweep the $B_{in}$ while applying a constant bias current and measure the voltage change with a Keithley 2400 Sourcemeter. We fabricated various devices consisting of Py/h-BN/Co and Co/h-BN/Co magnetic tunnel structures on chip scale suing such CVD h-BN.

## 3. Results and Discussion

Figure 2a shows the optical micrograph of large area CVD h-BN transferred on the $SiO_2$/Si substrate. Atomic force microscope (AFM) measurements reveal an effective thickness of ~5 Å for the h-BN layer on the $SiO_2$/Si substrate as shown in Fig 2 (b), which corresponds to single layer h-BN [30]. The Raman shift for the CVD h-BN has been found to be at ~1369 $cm^{-1}$ (Fig. 2(c)), which matches with the results of exfoliated h-BN [32]. We find the tunnel resistances of the h-BN contacts are in the range of 0.5 – 1 $kΩ.μm^2$ in our magnetic tunnel structures. The detailed tunneling characteristics of CVD h-BN barriers have been presented in our previous work, where an effective barrier height of ϕ ~ 1.49 eV was extracted [30].

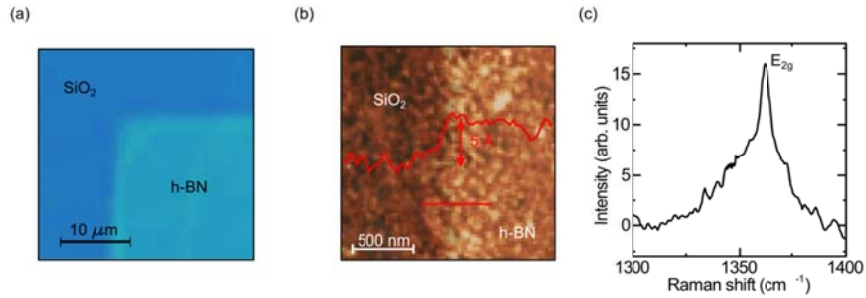

*Figure 2| CVD Hexagonal boron nitride: (a)* Optical micrograph of CVD h-BN on a $SiO_2$/Si substrate. *(b)* Atomic force microscopy image of a CVD h-BN layer on a $SiO_2$/Si substrate. The red line is the step scan showing an effective h-BN height of ~5 Å. *(c)* The Raman peak of CVD h-BN on a $SiO_2$/Si substrate.

Tunnel magnetoresistance measurements were performed in a four probe cross bar geometry (Fig. 3(a)), which enables the measurement of the tunnel junction resistance, while avoiding other effects such as anisotropic magnetoresistance of the FMs. We applied a fixed bias current between the top and bottom ferromagnetic electrodes of the junction while the voltage drop across the junction was measured as a function of the external in-plane magnetic field (B). A magnetization reversal of the FM electrodes occurs at fields corresponding to their respective coercivities. The switching field is achieved by using different materials and also through different geometries of the FM electrodes. Hence, their magnetizations can be aligned either parallel or antiparallel and the system enables us to observe two well-defined different resistance states due to difference between the threshold switching fields of the two FM electrodes.

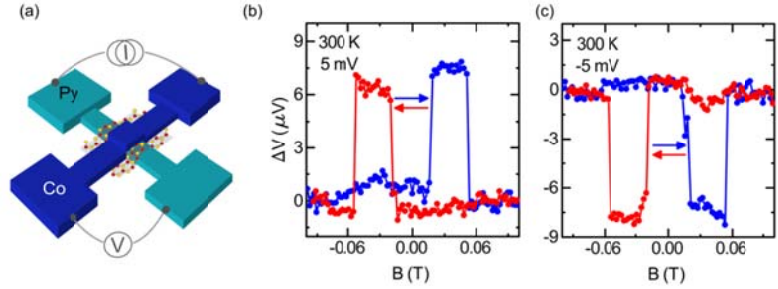

*Figure 3| TMR with Py/h-BN/Co MTJ at room temperature. (a)* Schematics of a Py/h-BN/Co MTJ in a four probe cross bar geometry. *(b, c)* TMR measurements with in-plane magnetic field B in Py/h-BN/Co MTJ for applied bias voltages of +/- 5 mV at 300 K. The arrows indicate the up and down B field sweep directions.

The TMR measurement in Fig. 3b clearly shows that the tunnel resistance abruptly increases upon switching the magnetization of the FM electrodes from a parallel to antiparallel magnetization configuration. The sharp switching of the FM electrodes and the flat region for the anti-parallel magnetization configuration is a strong indication of TMR [33-37]. The TMR ~0.15 % was calculated by using the relation TMR = $\frac{R_{AP}-R_P}{R_P}$ ×100 %, where $R_{AP}$ and $R_P$ are the tunnel resistances measured at antiparallel and parallel magnetization configuration of the two FM electrodes respectively [33]. As expected, the TMR signal also changes sign with reversing bias polarity (shown in Fig. 3c) [33-37]. The observation of TMR with h-BN barriers were reproduced on different devices and also in junction with different FM contacts. Figure 4 shows a TMR signal of 0.5% for a Co/h-BN/Co MTJ at 300 K, where different widths of Co electrodes were chosen to achieve different switching fields. These results suggest that one atomic layer of h-BN is sufficient enough to decouple the top and bottom FM layers, which lead to observation of TMR signal at room temperature [33-37].

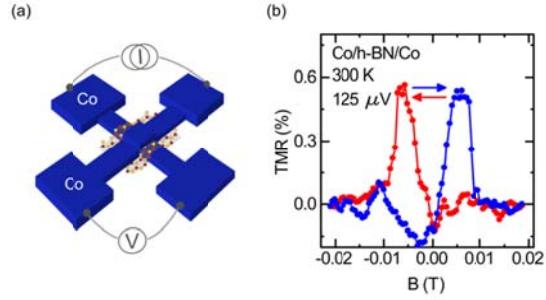

*Figure 4| TMR with Co/h-BN/Co MTJ at room temperature. (a)* Schematics of a Co/h-BN/Co MTJ in a four probe cross bar geometry. *(b)* TMR measurements with in-plane magnetic field in Co/h-BN/Co MTJ for applied bias voltages of 125µV at 300 K. The arrows indicate the up and down B field sweep directions.

The observed TMR is a consequence of spin-dependent tunneling from a FM through the h-BN barrier. The relationship between the observed TMR and tunnel spin polarizations (TSP) of the FM/h-BN interface can be explained by the Julliere model [33]. The spin polarization of the FM/h-BN contacts are estimated to be $P_1 \sim P_2 = P \sim 0.05 - 0.25$ % considering the relation TMR = $2P_1P_2/(1 - P_1P_2)$, where $P_1$ and $P_2$ are the tunnel spin polarizations of the Py/h-BN and Co/h-BN interfaces are assumed to be similar, respectively [33]. These values are less than previously reported TSP of P ~ 14 % in Co/h-BN contacts for spin injection into graphene in lateral spin transport devices, where FM electrodes are placed on the top of the h-BN/graphene channel [30]. Such spin polarization obtained in graphene spin transport devices are comparable to that obtained using $Al_2O_3$ and $TiO_2$ tunnel barriers [43]. Much higher values for spin polarization were predicted theoretically considering lattice matched FM/h-BN/FM vertical MTJ structures [26 -28]. We attribute the lower value obtained for our vertical MTJ devices to the possible contamination of the bottom Py or Co surface during the air exposure and h-BN transfer process [38-41]. The oxidation or contamination of the interface is known to produce strong spin-scattering that reduces the tunneling spin polarization even with conventional metal oxide tunnel barriers such as $Al_2O_3$ and MgO [38-41]. Higher spin polarization can be expected in structures grown in-situ without any air exposure. Therefore, the direct growth of h-BN on FMs and in-situ preparation of vertical MTJ layer stack is required to demonstrate the true spintronics potential.

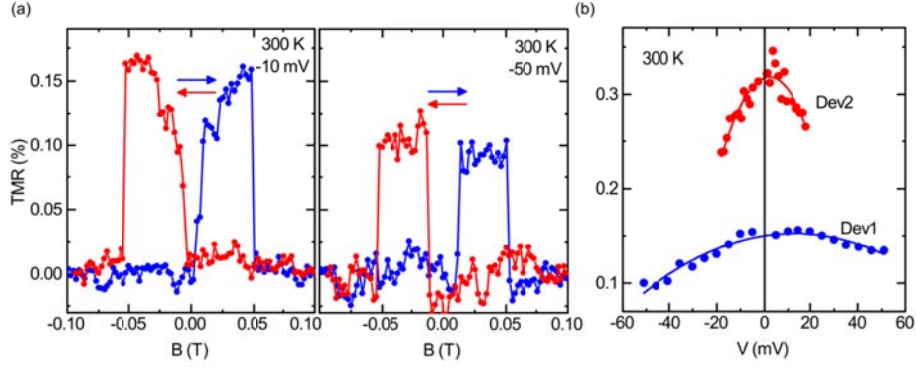

*Figure 5| Bias dependence of TMR with h-BN barrier at room temperature. (a) TMR plots for Dev1 at bias voltage of -10 and -50 mV at 300 K. The arrows indicate the up and down B field sweep directions. (b) Full bias dependence of TMR signal for two different Py/h-BN/Co devices measured at 300 K (bule circles - Dev 1, and red circles - Dev2). The lines are guide to the eye.*

Figure 5 shows the bias dependence of TMR for two devices with Py/h-BN/Co MTJs at room temperature. The TMR data of the Dev1 at -10 mV and -50 mV are shown in Fig. 5a and the complete bias dependence of TMR signal for both devices are plotted with bias voltage in Fig 5b. To be noted we restrict our measurements to low bias voltages < 50 mV, because of atomically thin h-BN tunnel barriers. The low voltage study is also appropriate considering the fact that tunneling dominates the conduction mechanism [35-37, 40]. The negative and positive bias range corresponds to spins tunneling through h-BN barrier from Py → Co and Co → Py respectively. We observe a decreasing trend of TMR at higher bias voltages with a maximum around zero voltage, which is consistent with behavior generally observed in spin polarized tunneling [35-41]. Our results reflect the typical behavior for MTJs, where the bias dependence of TMR relies upon the quality and height of the h-BN barrier, electronic properties and magnon excitations at the FM/h-BN interfaces [35, 40]. The observations of such characteristics also exclude any artifacts such as anisotropic magnetoresistance. The observed asymmetry in TMR with bias can be due to interfaces oxidation, different work functions, and densities of states for the two different FM electrodes [35-42]. The difference in the observed TMR values for two different devices can be due to slightly different interface conditions, which is not known exactly. The CVD grown h-BN is having mostly monolayer thickness, with few patches of bi-layer and thicker layers [30, 31]. Using these different thicknesses in lateral graphene spin transport devices, an enhancement in tunnel spin polarization for higher contact resistances corresponding to the thickness of h-BN barrier has been observed previously [30]. It would be interesting to investigate such thickness dependence of TMR in h-BN based MTJ structures and its proposed spin filtering attributes [26 -28].

## 4. Conclusions

In conclusion, we have demonstrated the feasibility of spin-dependent tunneling employing single-layer insulating h-BN barriers in MTJs. We measured a TMR effect of 0.3 – 0.5 % at room temperature corresponding to tunnel spin polarization P of 0.05 – 0.25 % for FM/h-BN junctions. The lower values of TMR and P are attributed to oxidation and contamination of the bottom FM electrode during h-BN transfer process. It is expected that much higher TMR ratios can be obtained through improvements in fabrication process with cleaner interfaces, incorporating multi-layer h-BN barriers [26-28], and use of h-BN/graphene heterostructures to achieve spin filtering [26]. These results demonstrate that uses of atomically thin large area CVD h-BN tunnel barrier are generic for a large class of devices requiring tunneling of spin polarized carriers and particularly interesting for technologically important magnetic tunnel junctions. Such h-BN tunnel barriers can also be employed for efficient spin injection into silicon and other semiconductor materials [44, 45]. Opportunities are growing with improvements in the growth process of large area CVD h-BN and h-BN/graphene van der Waals heterostructures on ferromagnets, which can be used in future spintronic devices to achieve desired contact resistances and large magnetoresistance signals [46-49].

**Acknowledgement**

The authors acknowledge the support from colleagues of Quantum Device Physics Laboratory and Nanofabrication Laboratory at Chalmers University of Technology. The authors would like to thank Jie Sun and Niclas Lindvall for sharing the recipe for 2D layer transfer process. This project is financially supported by the Nano Area of the Advance program at Chalmers University of Technology, EU FP7 Marie Curie Career Integration grant and the Swedish Research Council (VR) Young Researchers Grant. RSP acknowledges the financial support from the Department of Science and Technology, Government of India through nanomisson project (grant No. SR/NM/NS-1002/2010).